\documentclass{sig-alternate-05-2015}
\usepackage{comment}

\begin{document}

\setcopyright{acmcopyright}



\conferenceinfo{ODD}{v5.0: {ACM SIGKDD} 2018 Workshop August 2018, London UK}

%
\conferenceinfo{ODD}{v5.0: {ACM SIGKDD} 2018 Workshop August 2018, London UK}

\title{Grey-box Process Control Mining for Anomaly Monitoring and Deconstruction}
%
%
%
%
%

\numberofauthors{4} 
%
\author{
%
%
%
\alignauthor Andr\'es Vargas
\\
       \affaddr{Rensselaer Polytechnic Institute}\\
       \affaddr{Troy, NY USA}\\
       \email{vargaa5@rpi.edu}
\alignauthor MD Ridwan Al Iqbal\\
       \affaddr{Rensselaer Polytechnic Institute}\\
       \affaddr{Troy, NY USA}\\
       \email{iqbalm@rpi.edu}
\alignauthor John S. Erickson\\
       \affaddr{Rensselaer Polytechnic Institute}\\
       \affaddr{Troy, NY USA}\\
       \email{erickj4@rpi.edu}
\and  
\alignauthor Kristin P. Bennett\\
       \affaddr{Rensselaer Polytechnic Institute}\\
       \affaddr{Troy, NY USA}\\
       \email{bennek@rpi.edu}
}


\maketitle
\begin{abstract}
We present a new ``grey-box" approach to anomaly detection in smart manufacturing.   The approach is designed for tools run by control systems which execute recipe steps to produce semiconductor wafers.  Multiple streaming sensors capture trace data to guide the control systems and for quality control.  These controls systems are typically PI-controllers which can be modeled as an ordinary differential equation (ODE) coupled with a control equation, capturing the physics of the process.   The ODE ``white-box" models capture physical causal relationships that can be used in simulations to determine how the process will react to changes in control parameters, but they have limited utility for anomaly detection.    Many ``black-box" approaches exist for anomaly detection in manufacturing, but they typically do not exploit the underlying process control.   The proposed  ``grey-box" approach uses the process-control ODE model to derive a parametric function of sensor data.   Bayesian regression is used to fit the parameters of these functions to form characteristic ``shape signatures".  The probabilistic model provides a natural anomaly score for each wafer, which captures poor control and strange shape signatures. The anomaly score can be deconstructed into its constituent parts in order to identify which parameters are contributing to anomalies. 
We demonstrate how the anomaly scores can be used to monitor complex multi-step manufacturing processes to detect anomalies and changes and show how the shape signatures can provide insight into the underlying sources of  process variation that are not readily apparent in the sensor data.
\end{abstract}

%

%
%

%
%
\printccsdesc

\section{Introduction} \label{intro}

In semiconductor manufacturing, wafers are transformed into fully functional integrated circuits by passing each wafer through a sequence of tools or chambers to perform a series of deposition and etching processes.   Deposition is achieved via ``recipes" composed of a series of steps, each guided by control systems. Sensors continuously collect physical measurements, such as temperature and pressure, during each step.  
The resulting sensor traces are  used to guide the control system. Our approach strives to continuously 
monitor the sensors to perform quality control.

\begin{figure}[h!]
\caption{Temperature trace data for wafers (dots) through time and their associated Shape Signatures (colored curves). Each color represents a different wafer.    The black curve is the normal model (see latter sections for further discussion)}
\includegraphics[height=5cm,width=8cm]{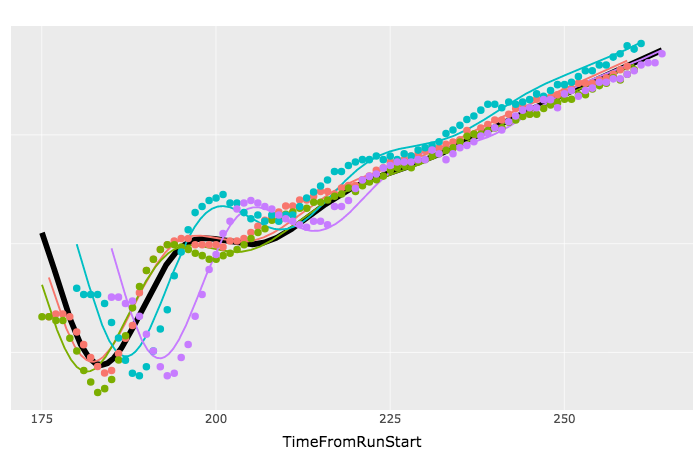}
\label{sampledat}
\end{figure}

Mathematically,  each  unique ($tool$, $sensor$, $step$, $wafer$) quadruple induces a time series $\{t,v_j(t)\}_{t \in t^{i,k}_l}$, in which $i$ indexes the tool, $j$ indexes the sensor, $k$ indexes the step, $l$ indexes the wafer, and $t^{i,k}_l$ is a vector containing the time points for tool $i$, step $k$, and wafer $l$.  The $t^{i,k}_l$  may have different numbers of time points, as well as a different sampling frequencies.  Figure \ref{sampledat} shows the raw sensor data from a particular $(tool, sensor, step, wafer)$ quadruple plotted through time, and colored by wafer.  It is readily apparent from this figure that an appropriate data model would be a damped harmonic oscillator with a perhaps linear driving force; we justify this assertion by fitting a damped harmonic oscillators to the data (colored curves).  In fact, all quadruples in our data can be modeled by a damped linearly driven oscillator as a consequence of the underlying PI control system; we demonstrate this fact mathematically in section \ref{PI control}.

The primary contribution of this work is using the vector of oscillator parameters, which we have called the \textit{shape signature}, corresponding to each quadruple as a summary statistic for the time series induced by the quadruple.  This serves as a powerful  data compression technique, as each time series, often with more than 100 observations, is reduced to just seven parameters.  More importantly, it transforms the raw data into a more information-rich parametric space.  For example, the parameter $\gamma$, i.e. the damping coefficient, encodes how fast the time series settles to the target set point; this could not have been determined easily from the raw time series.  In section \ref{mining}, we fit the shape signatures using constrained nonlinear bayesian regression. In section \ref{anom}, we develop an anomaly scoring function based on the objective function of the regression optimization procedure. Finally, in section \ref{monitor}, we present case studies of two different $(tool, sensor, step)$ triples, demonstrating how their respective anomaly scores can be monitored through time and deconstructed into the contributions from each shape signature parameter using a novel approach.  Since these parameters were shown to be analytically related to the underlying control system in section \ref{PI control}, such a deconstruction is directly relevant to process engineers.

Our data set consists of about four months of data from two different ``recipes", containing trace data from 16 sensors of six types: current, voltage, temperature, power, pressure, and the angular position of a throttle valve.  One of the monitored recipes contained eleven steps and was measured on six different tools; the other recipe is fourteen steps and was measured on seven tools; so in total there are $16 \times (11 \times 6 + 14 \times 7)=2624$ triples.  Since there are between 50 and 300 wafers in each triple, there are at least $2624 \times 50 = 131,200$ quadruples that must be fitted with shape signatures.  We used distributed computing to greatly minimize algorithm run time; succesfully fitting each quadruple with a characteristic shape signature.



Feature construction for anomaly detection in semiconductor manufacturing has been studied extensively. A common approach uses dynamic time warping or interpolation to massage the sensor data into a common time framework. General-purpose anomaly detection algorithms are then applied, with variants of PCA and multiway analysis being popular choices \cite{wise1999comparison,cherry2006multiblock,ge2010semiconductor}. Most approaches treat sensor data as single (long) traces instead of subdividing into recipe steps consistent with the control algorithms. Unlike our shape signature method, these approaches do not leverage knowledge of the underlying recipes and controllers. Shape signatures are a powerful new representation that could be incorporated into any outlier or anomaly detection algorithm \cite{aggarwaloutlieranalysis} and can indicate which steps, sensors, and parameters of the control systems are associated with each anomaly.  

An approach similar to our shape signature method constructs ten features, some related to the parameters of the damped harmonic oscillator, including amplitude and settling time \cite{haq2016feature}. Improving on this, our approach constructs ``grey-box" features that, along with independently encoding semantically meaningful information and having the ability to be monitored and incorporated into an anomaly detection framework, provide an intricate picture of the underlying physics when considered jointly as a damped harmonic oscillator.

\section{Grey-box Model} \label{PI control}

In this section, we show how the underlying ``white-box" model for PI controllers leads to the classical equation of damped harmonic motion that forms the basis of our ``grey-box" approach \cite{bequetteprocesscontrol}.
\subsection{Process Model}
In classical control theory a PI Controller controls an output variable $v(t)$ (e.g. temperature) by controlling some input variable $u(t)$ (e.g. power).  We assume that the process model relationship between $v$ and $u$ can be modeled by a first order differential equation:
\begin{align} \label{eq1}
v'(t) = -\frac{1}{\tau_p} v(t) + \frac{k_p}{\tau_p} u(t)\ ,
\end{align}
where $\tau_p$ is the process time constant and $k_p$ is the process gain.  We assume that $\tau_p$ and $k_p$ are nonzero.  

For the input and output variables $u$ and $v$, the process transfer function $g_p(s)$ must satisfy the relationship $V(s) = g_p(s) U(s)$, where $V$ and $U$ are the Laplace transforms of $v$ and $u$, respectively; and $s$ is a Laplace domain variable.  We can derive $g_p$ by taking the Laplace transform of both sides of (\ref{eq1}):

$$sV(s) - v(0) = -\frac{1}{\tau_p} V(s) + \frac{k_p}{\tau_p} U(s)$$
Assuming the initial condition $v(0)=0$, we obtain 
$$V(s) = \frac{k_p}{\tau_p s +1}U(s)\ .$$
So, $g_p(s) = \frac{k_p}{\tau_p s +1}$.

\subsection{Control Model}
A PI controller specifies $u$ as function of the error between an observed value of $v$ and the desired value:

\begin{align} \label{eq2}
u(t) = u_0 + k_c e(t) + \frac{k_c}{\tau_I} \int_0^t e(x) dx\ ,
\end{align}
where $u_0$ is a bias term, $k_c$ is the proportional gain, $\tau_I$ is the integral time, and $e(t) = r(t) - v(t)$ is the error.  Note, $r(t)$ is the desired temperature or ``set point" at time $t$.  We assume that $k_c$ and $\tau_I$ are nonzero.

The controller transfer function $g_c(s)$ must satisfy the relationship $U(s) = g_c(s) E(s)$.  Taking the Laplace transform of both sides of (\ref{eq2}), we obtain

$$U(s) = \frac{u_0}{s} + k_c E(s) + \frac{k_c}{\tau_I} \frac{E(s)}{s} .$$\\

Assume the bias $u_0$ is zero and manipulate to obtain
$$U(s) = \frac{k_c(\tau_Is + 1)}{\tau_Is} E(s)\ .$$
So, $g_c(s) = \frac{k_c(\tau_Is + 1)}{\tau_Is}$.

\subsection{Closed-Loop Model}
To obtain the closed loop model (transfer function), we couple the process model and the control model.  First, recall the process relationship $V(s) = g_p(s) U(s)$.  Substituting in the controller relationship $U(s) = g_c(s)E(s)$, we have $V(s) = g_p(s)g_c(s)E(s)$.  Also, since $e(t) = r(t)-v(t)$, $E(s) = R(s) - V(s)$, and so 
$$V(s) = g_p(s)g_c(s)(R(s) - V(s))\ .$$ 
Solve for $V(s)$ to obtain 
$$V(s) = \frac{g_p(s)g_c(s)}{1 + g_p(s)g_c(s)}R(s)\ ,$$
thus the closed loop transfer function is 
 \begin{align} \label{closed loop trans}
 g_{CL}(s) = \frac{g_p(s)g_c(s)}{1 + g_p(s)g_c(s)}\ .
\end{align}
We can write (\ref{closed loop trans}) as 
\begin{align}
g_{CL}(s) =\frac{\tau_Is + 1}{\frac{\tau_I\tau_p}{k_pk_c}s^2 + \frac{\tau_I(1+k_pk_c)}{k_pk_c} s + 1} = \frac{\tau_Is + 1}{ms^2 + \gamma s + k}\ ,
\end{align}
where $m=:\frac{\tau_I\tau_p}{k_pk_c}$, $\gamma:=\frac{\tau_I(1+k_pk_c)}{k_pk_c}$, and $k:=1$.
\subsection{Recovering Damped Linearly Driven Harmonic Motion}
We now have 
\begin{align}
V(s) &= g_{CL}(s)R(s) = \frac{\tau_Is + 1}{ms^2 + \gamma s + k}R(s) \\
& \Rightarrow (m s^2 + \gamma s + k) V(s) = (\tau_I s + 1) R(s)\ .
\end{align}
Now use the inverse Laplace transform to bring both sides back to the time domain, again assuming that the relevant initial conditions are zero:
\begin{align} \label{damped harmonic motion 1}
m v''(t) + \gamma v'(t) + kv(t) = \tau_I r'(t) + r(t)\ .
\end{align}
Observe that (\ref{damped harmonic motion 1}) is the second order linear differential equation for damped harmonic motion.  For spring motion, $m$ is the mass of the object attached to the spring, $\gamma$ is the damping coefficient, $k$ is the spring constant, and $\tau_I r'(t) + r(t)$ is the driving force.  We divide through by $m$ and express everything in terms of the underlying parameters to obtain 

\begin{align}  \label{damped harmonic motion two}
v''(t) + \frac{1+k_pk_c}{\tau_p} v'(t) + \frac{k_pk_c}{\tau_I\tau_p} v(t) = \frac{k_pk_c}{\tau_p}r'(t) + \frac{k_pk_c}{\tau_I\tau_p}r(t)\ .
\end{align}

Note that equation (\ref{damped harmonic motion two}) can also be recovered for the input variable $u$.  Explicitly, we have the Laplace domain relationship
\begin{align} \label{input eq}
U(s) &= g_{c}(s) E(s) = g_c (R(s)-V(s)) \nonumber \\
&= g_c(s) (R(s)-g_p(s)U(s)) \nonumber \\
& = \frac{k_c(\tau_Is + 1)}{\tau_Is} (R(s)-\frac{k_p}{\tau_p s +1}U(s))
\end{align}
We can manipulate (\ref{input eq}) to obtain
\begin{align*}
&[\tau_I\tau_p s^2 + \tau_I(k_pk_c+1)s + k_pk_c]U(s) = [k_c\tau_I\tau_ps^2 + k_c(\tau_p+\tau_I)s \\
&+k_c]R(s)\ . 
\end{align*}
Taking the inverse Laplace transform of both sides and assuming the relevant initial conditions are zero, we have
\begin{align}  \label{damped harmonic motion three}
&u''(t) + \frac{1+k_pk_c}{\tau_p} u'(t) + \frac{k_pk_c}{\tau_I\tau_p} u(t) \nonumber \\
&= k_cr''(t) + \frac{k_c(\tau_p+\tau_I)}{\tau_I\tau_p}r'(t) + \frac{k_c}{\tau_I\tau_p}r(t)\ .
\end{align}
Note that (\ref{damped harmonic motion three})  differs from (\ref{damped harmonic motion two}) only in the right hand side.  

Assume $r(t)$ is linear, say $r(t):=q_1t+q_2$, then we can write the right hand side of (\ref{damped harmonic motion two}) as $at+b$, where $b=\frac{k_pk_c}{\tau_p}(q_1+\frac{q_2}{\tau_I})$ and $a=\frac{k_pk_cq_1}{\tau_I\tau_p}$.  We can also write the right hand side of $(\ref{damped harmonic motion three})$ as $at+b$, only now $b=\frac{k_c}{\tau_I\tau_p}[q_1(\tau_p+\tau_I)+q_2]$ and $a=\frac{k_cq_1}{\tau_I\tau_p}$.  In either case, we can redefine $\gamma$ to be $\frac{1+k_pk_c}{\tau_p}$ and $k$ to be $\frac{k_pk_c}{\tau_I\tau_p}$, to obtain an equation of the form  
$$v''(t) + \gamma v'(t) + kv(t) = at+b\ .$$
The solution to the above differential equation is
\begin{align} \label{diff sol}
v(t) = Re^{-\gamma t/2}cos(\omega t - \phi)  + ct + y\ ,
\end{align}
where $c= a/k$, $y = (b-c\gamma)/k$,  $\omega:=\sqrt{4k - \gamma^2}/2 \in \mathds{R}$, $R:=A/\text{cos}\phi = B/\text{sin}\phi$, and $A$ and $B$ are constants that can be determined from the initial conditions.  See \cite{elementarydiffeqs} for a derivation of this solution.  The oscillator (\ref{diff sol}) is a very well mathematically understood object and its parameters have clear physical interpretations: $R$ is the amplitude, $\gamma$ is the damping coefficient, $\omega$ is the frequency, $\phi$ is the phase shift, $c$ is the slope, and $y$ is the vertical shift. 

We have shown that PI control leads to (\ref{diff sol}) for both the input and output variables of the controller, and the only differences between the input and output variables are the slope ($a$) and intercept ($b$) of the applied force.  In addition, we know that PI control is used at every step of every tool.  Recall that the sensors available are temperature, power, voltage, current, pressure, and throttle valve; and also note that temperature is controlled as a function of power, voltage is controlled as a function of current, and pressure is controlled as a function of throttle valve.  Thus, (\ref{diff sol}) is an appropriate model for the time series induced by every $(tool,sensor,step,wafer)$ quadruple in our data.  In section \ref{mining}, we estimate the parameters of (\ref{diff sol}) for each quadruple; we call these parameters ``shape signatures".

Suppose the parameters of (\ref{diff sol}) have already been estimated.  Then it is easy to reverse engineer the ODE parameters $a$, $b$, $k$, and $\gamma$ using the equations immediately following (\ref{diff sol}). Now consider  the output variable formulation  using $v$ as the sensor.  We we have shown the following analytical relationships between the ODE parameters and the underlying control parameters: 
\begin{align} \label{reverse engineer}
\begin{cases} 
\gamma = \frac{(1+k_p k_c)}{\tau_p} \\
k = \frac{k_p k_c}{\tau_I \tau_p}\\
a = \frac{k_pk_cq_1}{\tau_I\tau_p}  \\
b = \frac{k_pk_cq_1}{\tau_p} + \frac{k_pk_cq_2}{\tau_I\tau_p}
\end{cases}
\end{align}
Using the above, we can understand the behavior of the oscillator in terms of the control system parameters.  For instance, we know that (\ref{diff sol}) oscillates if $\omega \neq 0 \Leftrightarrow \omega^2>0 \Leftrightarrow \gamma^2<4k$.  By substitution, we obtain the condition in terms of the control system parameters:
$$\tau_I < \frac{4k_pk_c\tau_p}{(1+k_pk_c)^2}\ .$$  

In addition, the first two equations of (\ref{reverse engineer}) imply that $\tau_p = \frac{1}{\gamma -\tau_Ik}$ and $k_p = \frac{k\tau_I}{k_c(\gamma-\tau_Ik)}$.  $\tau_I$ and $k_c$ are often known in practice because they will often have been tuned to optimize certain domain-specific criteria.  If they are known, then we can analytically solve for the process parameters $\tau_p$ and $k_p$, which are never known in practice.  We also get $r(t)$ for free if the controller parameters are known: $q_1=a/k$ and $q_2=\frac{b-\tau_Ia}{k}$.

If the controller parameters are not known, then the first two equations imply $\tau_I = \frac{\gamma \tau_p-1}{k\tau_p}$ and the first equation can be rewritten as $\bar k = \gamma \tau_p-1$, where $\bar k = k_p k_c$.  Again, we have $q_1=a/k$ and $q_2=\frac{b-\tau_Ia}{k}$.  Thus, we have expressed the parameters $\tau_I$, $\bar k$, $q_2$, and $q_1$ as functions of  the single parameter $\tau_p$.  In other words, we have reduced the entire control system to a single degree of freedom $\tau_p$. This makes it easy to see how $\tau_p$ affects the behavior of the system.  For instance, $\bar k > 0$ is a necessary and sufficient condition for stability of PI control \footnote{This condition is necessary and sufficient only for the first order process assumed in (\ref{eq1})}, which from our analysis is equivalent to the condition $\tau_p > 1/\gamma$.  Also, observe that $\lim_{\tau_p \rightarrow \infty} \tau_I(\tau_p) = \gamma/k$, so for large enough $\tau_p$, $\tau_I \approx \gamma/k$, and hence $q_2 \approx \frac{bk-\gamma a}{k^2}$.

Our approach implicitly represents the control system parameters.  It is not ``white box", since it only provides full recovery of the underlying closed loop parameters in some cases.  It is also not ``black box", since the damped harmonic oscillator parameters (later fitted through bayesian regression) are intimately connected to the underlying control system.  Thus, it is ``grey-box" and as such can be uniquely used to generate insights to guide manufacturing processes.

\subsection{Empirical Evidence} \label{emp ev}
\begin{figure}[h!]
\caption{Temperature Trace data, colored by wafer, follows damped linearly driven harmonic motion.}
\includegraphics[height=4cm]{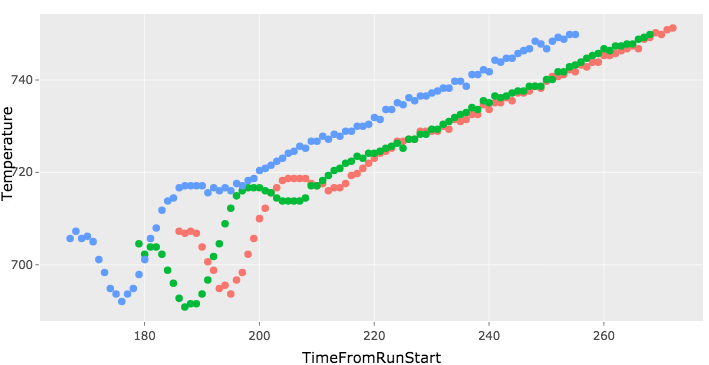}
\label{data_examp_dld}
\end{figure}

\begin{figure}[h!]
\caption{Voltage Trace data, colored by wafer, follows damped linearly driven harmonic motion.}
\includegraphics[height=4cm]{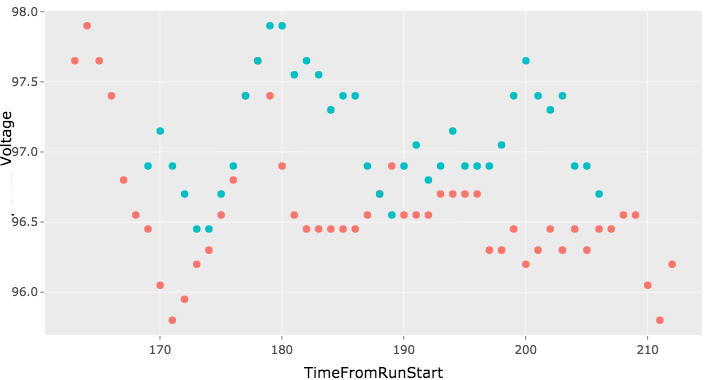}
\label{data_examp_dld2}
\end{figure}

\begin{figure}[h!]
\caption{Voltage Trace data, colored by wafer, follows damped constantly driven harmonic motion.}
\includegraphics[height=4cm]
{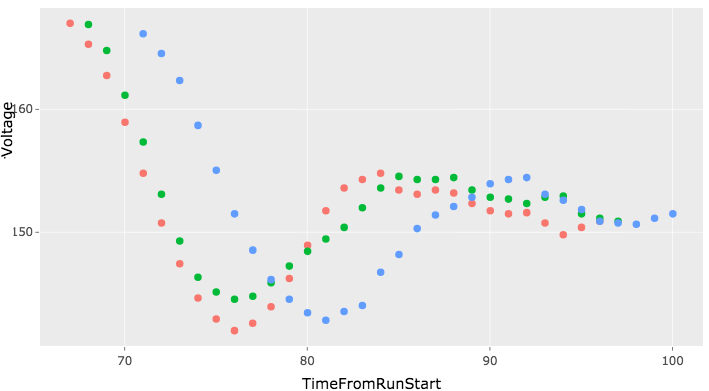}
\label{data_examp_dcd}
\end{figure}

\begin{figure}[h!]
\caption{Current Trace data, colored by wafer, follows exponential motion.}
\includegraphics[height=4cm]{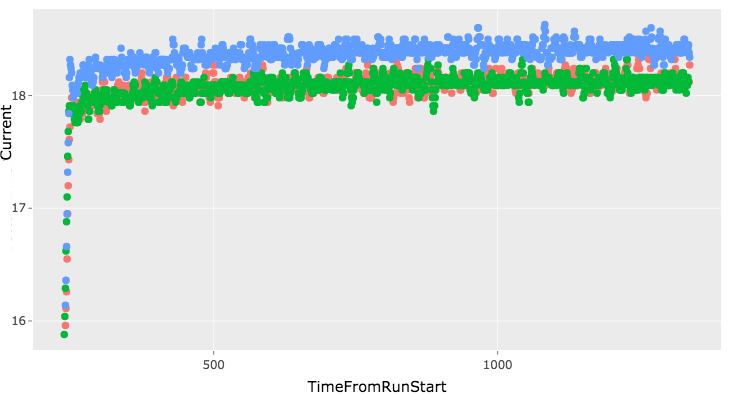}
\label{data_examp_e}
\end{figure}

\begin{figure}[h!]
\caption{Throttle Valve Trace data, colored by wafer, follows constant motion.}
\includegraphics[height=4cm]{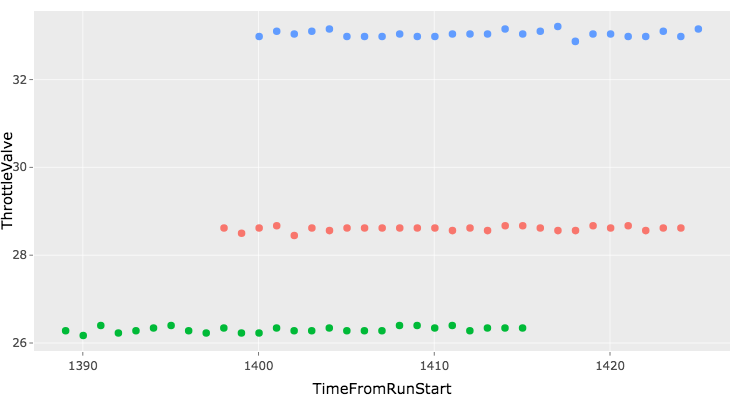}
\label{data_examp_c}
\end{figure}

Figures \ref{data_examp_dld}-\ref{data_examp_c} present raw data from a sample of $(tool,$ $sensor,$ $step,$ $wafer)$ quadruples.  In all cases, the damped linearly driven harmonic oscillator (\ref{diff sol}) captures the dynamics, thus agreeing with the previous theoretical derivation.  Importantly, figures \ref{data_examp_dcd}-\ref{data_examp_c} show data that appears to follow constantly driven harmonic, exponential, and constant motion; which can be parametrized as $v(t) = Re^{-\gamma t/2}cos(\omega t - \phi)  + y$, $v(t) = Re^{-\gamma t/2} + y$, and $v(t)=y$; respectively.  These equations are special cases of (\ref{diff sol}) when certain parameters are set to zero.  Thus, our claim that all quadruples follow damped linearly driven harmonic motion holds.

\section{Mining Damped Harmonic Oscillators} \label{mining}
Equation (\ref{diff sol}) gives the explicit form of the mathematical relationship between the sensor $v$ and time $t$ in the time series induced by each $(tool, sensor, step, wafer)$ quadruple.  To estimate the parameters $R, \gamma, \omega, \phi, c$, and $y$, we use Bayesian regression over sets of contiguous-in-time wafers. These sets are defined by the engineer to be lots of wafers processed together.  Without loss of generality, fix the tool, sensor, and step  so that the $i$, $j$, and $k$ indices are not necessary. Let $L = \{l_m\}_{m=1}^M$ be a set of contiguous-in-time wafers, let $t_{l_m}$ be the set of time points corresponding to wafer $l_m$, and let $z_m$ be the sensor values recorded. 
Define $S_m := (\gamma_m,R_m,\omega_m,y_m,\phi_m,c_m,x_m) \sim N(\alpha(\mu_S,diag(\sigma_S)^2)$ for all $m \in \{1,...,|L|\}$, where $\gamma_m$, $R_m$, $\omega_m$, $y_m$, $\phi_m$, and $c_m$ are the parameters of equation (\ref{diff sol}) corresponding to to wafer $l_m$, $x_m$ is a horizontal shift term that has been included to control for the fact that recipe steps often don't start at time 0 (we give a more detailed justification for the inclusion of $x_m$ in a few paragraphs),  $\sigma_S$ is a vector whose elements are the standard deviations of the respective elements of $S_m$ for all $m=1,...,|L|$, and $diag(a)$ is a diagonal matrix whose diagonal elements are the elements of the vector $a$.  Note that, since $S_m$ is multivariate Gaussian, the diagonal covariance implies independence of its individual elements.

We assume that the wafers are independent and that the sensor measurements exhibit independent Gaussian noise. 
Specifically, assume $S_i$ and  $S_j$  are independent for all $i \neq j$ and let $Z_m|S_m \sim N(\harpoon \alpha(t_{l_m},S_m),\sigma^2 I)$, where $\harpoon \alpha$ is defined, for any vector of parameters $s=(\gamma,R,\omega, y, \phi, c,x)$ and set of times for $N$ observations $t=\{t_1,t_2,...,t_{N}\}$, by 
\begin{equation}
\harpoon \alpha(t,s) :=
\begin{pmatrix}
\alpha(t_1,s)\\
\alpha(t_2,s) \\
\vdots \\
\alpha(t_N,s)
\end{pmatrix}  
\end{equation}  
with 
$$\alpha(t_i,s):=Re^{-\gamma (t_i-x) /2}cos(\omega (t_i-x) - \phi)  + c(t_i-x) + y\ .$$

Note that the sampling frequency and number of elements in $t_{l_m}$ varies with $m$.  The graceful handling of this issue is a major benefit of this formulation over prior work. 

\newdef{remark}{Remark}
 
We seek to find   $s_m$, $m=1,...,M$, and hyperparameters $\sigma$, $\mu_S$, and $\sigma_S$ that maximize the joint probability of the wafers and elements of $S_m$, for each $m$.  We assume that this joint probability can be decomposed as $$\prod_{m=1}^{|L|} P(z^m|s_m; \sigma) P(s_m; \mu_S, \sigma_S)\ ,$$ while enforcing the physical constraints $\omega_m \geq 0$ and $\phi_m \in [-\pi/2,\pi/2]$.  We equivalently minimize the negative log likelihood $-\sum_{m=1}^{|L|}\{ \text{ln}P(z^m | s_m;\sigma) + \text{ln}P(s_m;\mu_S,\sigma_S) \}$.  This can be formulated as the optimization problem

\begin{equation} \label{shapesig opt}
\begin{array}{ll} 
\min_{s_1,...,s_{|L|}, \sigma,\mu_S,\sigma_S} 
& 
-\sum_{m=1}^{|L|} \text{ln}P(z^m | s_m;\sigma)  \\
&-\sum_{m=1}^{|L|}\text{ln}P(s_m;\mu_S,\sigma_S) \\
\text{subject to} &
 \omega_m \geq 0, m=1,...,|L| \\
& \phi_m \in [-\pi/2,\pi/2], m=1,...,|L|
\end{array}
\end{equation}
One part of the objective involves the sum of square residuals between the actual and predicted sensor values 
\begin{equation}
\label{eq:error}
\begin{array}{ll}
-\text{ln}P(z_m | s_m;\sigma^2) := &
\frac{\text{ln}(\sigma^2)}{2} |t_{l_m}| +\\ &\frac{1}{2 \sigma^2}  \sum_{t \in t_{l_m}}(z_{m,t} - \alpha(t,s_m))^2. \\ 
\end{array}
\end{equation}
Defining $||\cdot||^2_A:={(\cdot)^T A (\cdot)}$, the other objective term involves the prior distribution of the shape signature parameters
\begin{equation}
\label{eq:prior}
\begin{array}{ll}
 -\text{ln}P(s_m;\mu_S,\sigma_S)  := & \sum_{d=1}^7 \text{ln}((\sigma_S)_d) +\\ 
& \frac{1}{2}  ||s_m - \mu_S||_{diag(\sigma_S)^{-1}}^2.  \\
\end{array}
\end{equation}


We solve (\ref{shapesig opt}) using block coordinate descent. First starting with initial guesses for the hyperparameters ($\sigma$, $\mu_S$, $\sigma_S$), we minimize  with respect to the $s_m$'s while holding the hyperparameters constant, and then we minimize with respect to the hyperparameters while holding all the $s_m$'s constant.  The Gaussian assumptions result in a closed from solution for the hyperparameter minimization, so only the minimization over the $s_m$'s requires an iterative procedure.  For the iterative procedure, we use modified Newton descent, via the R function ``nlminb" from the ``stats" package, because the dimension is small enough to make the analytical Hessian tractable to derive and invert.  Due to the independence of the $S_m$'s, the iterative optimization procedure can be performed for each $s_m$ independently.  

The parameter $x_m$ is not included in the iterative optimization procedure.  Instead, it is hard-coded to the minimum time in the corresponding induced time series to control for the fact that each induced time series corresponds to one of many steps in the deposition recipe, and so it need not start at 0.  The parameter $\phi_m$, often referred to as the phase shift, \textit{is} included in the iterative procedure in order to capture horizontal shift that is not merely an artifact of the recipe, but rather an important feature of the wafer.

\newdef{definition}{Definition}
\begin{definition} \label{shapesignature}
For a set of contiguous-in-time wafers $L=\{l_m\}_{m=1}^M$, denote the optimal value of $s_m$, from the optimization program (\ref{shapesig opt}), by $s_{l_m}$, $\forall m \in \{1,...,M\}$.  We call $s_{l_m}$ the \textit{shape signature} of wafer $l_m$.
\end{definition}

Note that (\ref{shapesig opt}) fits shape signatures of the set of contiguous-in-time wafers $L$ simultaneously. This is preferable to fitting one wafer at a time, since   some wafers have too few observations to optimize in isolation.  By learning wafers in sets, we share information between contiguous wafers enabling fits when sensor data is sparse, while also creating an estimate of the prior distribution of the shape signature parameters.   This stabilizes  and improves the quality of the estimates.  In the next section, we show that the prior also plays a crucial role in establishing the normal model of chamber operation.

Knowledge of the manufacturing process is used to choose $L$.  Specifically, on the factory floor, the wafers are processed in lots, with each lot of wafers in its own container, called a foup.  Robots ferry the foups to different chambers.  Since the chamber can only perform the deposition on one wafer at a time,  robotic arms lift the wafers for deposition into the chamber one after the other in succession.  Engineers expect all wafers in the foup to follow the same manufacturing process thus they share very similar shape signatures modulo measurement errors.  Therefore, the wafers in each lot satisfy the contiguous-in-time requirement.  Thus, we solve (\ref{shapesig opt}) for each lot of wafers $L$ in each $(tool,sensor,step)$ triple in order to obtain shape signatures for all the wafers in our data.\footnote{The choice of $L$ differs slightly during the \textit{initial fit}; see section \ref{anom}.}


We use the parallel computing capabilities of  R  to quickly solve $(\ref{shapesig opt}$) for each lot in each $(tool,sensor, step)$ triple.  There are 2624 triples, and between 50 and 300 wafers in each triple, so distributed computing results in an immense reduction in computation time.  We output the shape signatures for each $(tool,sensor,step,wafer)$ quadruple into one ``data.table"\footnote{A ``data.table" is a special R data structure that is essentially a matrix that is highly optimized for big data storage and manipulation.} per tool, where each column is a unique combination of sensor and step, and each row is a unique wafer.  We order the rows with respect to time so that we now have a single multivariate time series for each tool.


\subsection{Sample Shape Signature Fits}
\begin{figure}[h!]
\caption{Voltage shape signatures, colored by wafer, for damped constantly driven harmonic motion.  Slopes: red=-0.013, blue=-0.011.}
\includegraphics[height=4.5cm]{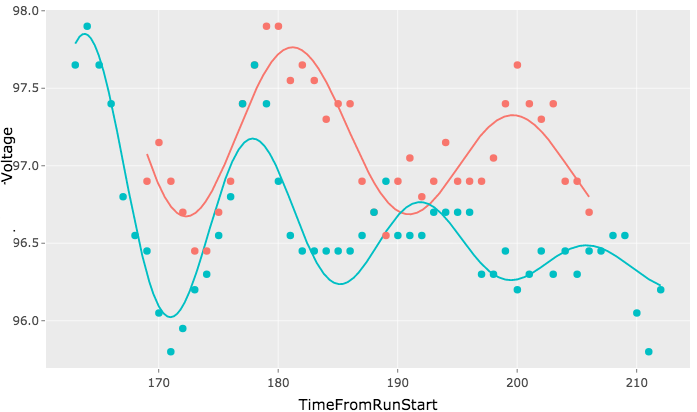}
\label{ss_examp_dcd1}
\end{figure}

\begin{figure}[h!]
\caption{Voltage shape signatures, colored by wafer, for damped constantly driven harmonic motion.  Slopes: green=0.104, blue=0.093, red=0.095.}
\includegraphics[height=4.5cm]{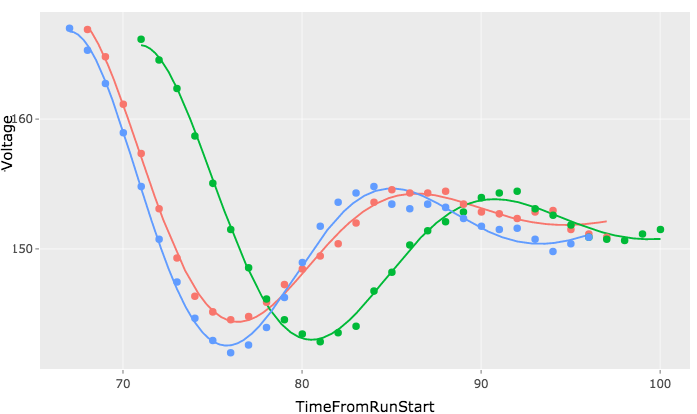}
\label{ss_examp_dcd2}
\end{figure}

\begin{figure}[h!]
\caption{Current shape signatures, colored by wafer, for exponential motion.  The raw data blocks view of the shape signatures, so it has been excluded in this image.  Damping coefficient ($\gamma$): red=0.347, blue=0.250, green=0.360.  Frequency ($\omega$):  red=0.168, blue=0.005, green =0.153,.}
\includegraphics[height=4.5cm]{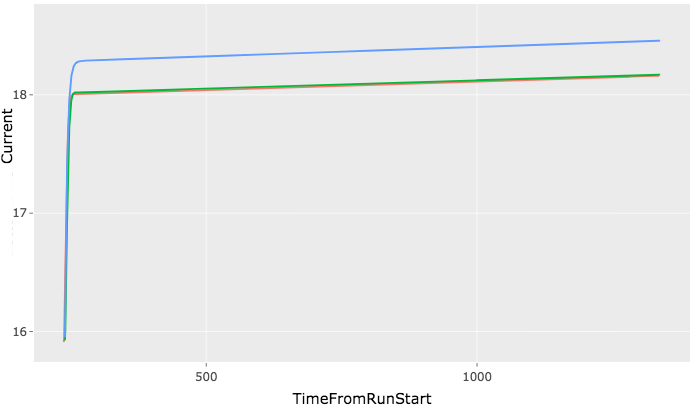}
\label{ss_examp_e}
\end{figure}

\begin{figure}[h!]
\caption{Throttle Valve shape signatures, colored by wafer, for constant motion.  Slopes: green=5e-04, blue=1e-04, red=6e-04.  Amplitudes ($R$):  green=-4.7e-04, blue=-2.4e-02, red=-1e-04.  Frequencies $(\omega$):  green=0.264, blue=0.344, red=0.438. Damping coefficient ($\gamma$): green=-0.522, blue=0.074, red-0.248.}
\includegraphics[height=4.5cm]{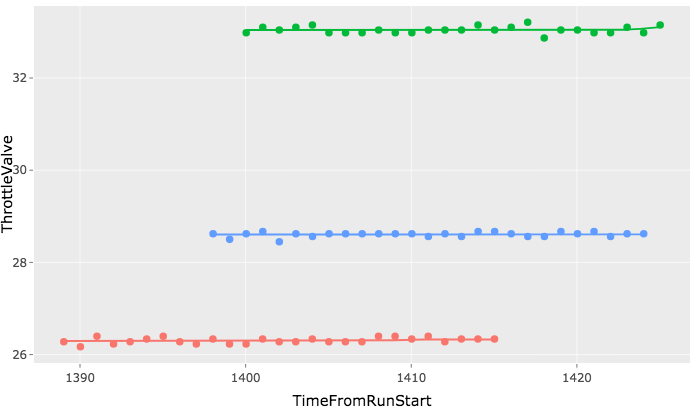}
\label{ss_examp_c}
\end{figure}

In this section we examine fitted shape signatures for the same trace data examples discussed in section \ref{emp ev}.  In Figures \ref{ss_examp_dcd1} and \ref{ss_examp_dcd2}, the algorithm estimates all slopes to be close to zero since the data does not exhibit a significant linear trend.  In Figure \ref{ss_examp_e}, the shape signatures do not exhibit oscillations, thus conforming to what we observed in figure \ref{data_examp_e}.  Note that oscillation decreases as the difference between the damping coefficient ($\gamma$) and the frequency ($\omega$) increases.  We can see this occurring in Figure \ref{ss_examp_e}; when the damping coefficient drops in the blue curve, so too does the frequency.  In Figure \ref{ss_examp_c}, the amplitudes and slopes are all small in magnitude, resulting in shape signatures representing constant functions.  Regularization or model selection can encourage parameters of the simpler models to be \textit{identically} zero.

\begin{figure}[h!]
\caption{Temperature shape signatures, colored by wafer, for damped linearly driven harmonic motion.  Slopes: green=0.595, blue=0.590, red=0.595.}
\includegraphics[height=4.5cm]{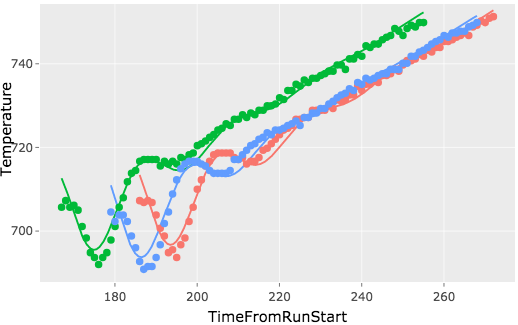}
\label{ss_examp_dld}
\end{figure}


\section{Anomaly Scoring} \label{anom}
An integral part of any anomaly detection algorithm is the computation of an anomaly score for each data point, such that a large anomaly score indicates the data does not represent an expected pattern and a low anomaly score indicates expected behavior.  This can be achieved by defining an anomaly scoring function that maps each data point to a real number that is high when the data point does not fit the expected pattern and low if it fits. Creating a viable anomaly score depends on the underlying assumptions of what is ``normal", what is ``not normal", and domain knowledge, making anomaly scoring more of an art than science \cite{aggarwaloutlieranalysis}.  In our case, we were given two requirements for our anomaly scoring function: 1. Penalize poor fit of shape signature to sensor data; 2.  Penalize large differences of a fitted shape signature from a ``normal" shape signature model. 

With these requirements in mind, consider the objective function of (\ref{shapesig opt}), with $|L|=1$:
\begin{align} 
-\text{ln}P(z_1|s_1;\sigma) - \text{ln}P(s_1;\mu_S,\sigma_S) , \label{anom score motivate}
\end{align}
and suppose $\sigma$, $\mu_S$, $\sigma_S$ are known.  Then (\ref{anom score motivate}) measures how aberrant $s_1$ is, where aberrant is defined as ``low probability of occurring" using the probability density function $P(s_1,z_1)=P(z_1|s_1)P(s_1)$.  Note that the first term of (\ref{anom score motivate}) encodes deviance with respect to the data since it is the negative  log of the likelihood of $z_1$. The second term measures deviance of $s_1$ with respect to the prior distribution.  Thus, encoding the normal model of operation into the prior $P(S;\mu_S,\sigma_S)$ would cause (\ref{anom score motivate}) to satisfy the requirements 1 and 2.  The following definition explains how to do this.

\begin{definition}
For a given $(tool,sensor,step)$, triple, take a set of contiguous-in-time wafers at the start of the data collection period.  Assume that the data contained in these wafers were generated by the same joint distribution \begin{equation}
P(S,Z;\sigma^*,\mu_S^*,\sigma_S^*) = P(Z|S;\sigma^*)P(S;\mu_S^*,\sigma_S^*).
\end{equation} The normal model is defined as ($\sigma^*, \mu_S^*,\sigma_S^*$). We can then solve $(\ref{shapesig opt})$, with $L$ now containing the afore-mentioned wafers, to estimate the normal model.   We refer to the above procedure of learning the normal model as the \textit{initial fit}. 
\label{Normal Model}
\end{definition}

Note that, unlike the procedure from the previous section, the wafers used in the initial fit are not required to be from the same lot.  In fact, it is better for them to span multiple lots, as this reduces variance of the normal model parameter estimates. On the other hand, using too many wafers is also detrimental, as it will bias the parameter estimates, since we would expect the underlying process to drift from where it began as time progresses.  In practice, we used four lots of wafers for the initial fit for each triple.  Based on the normal model, we define an anomaly scoring function which satisfies requirements 1 and 2:
\begin{definition}
Given the normal model $(\sigma^*, \mu_S^*,\sigma_S^*)$, the anomaly score of a shape signature $s$, with corresponding wafer $l$ and data $z=\{z_t\}_{t \in t_l}$, is 
\begin{equation}
\text{anom}(s) := -\text{ln}P(z|s; \sigma^*) - \text{ln}P(s;\mu_S^*, \sigma_S^*)\ .
\end{equation}
\end{definition}

We now have a way of assigning an anomaly score to each shape signature.  By construction, the anomaly score is a function of the shape signature parameters, which encode important information about the underlying control process that is hidden in the raw data.  Thus, we expect for these anomaly scores to be better for identifying and deconstructing anomalies than methods that only use the trace data without exploiting the underlying control process.

\section{Monitoring and Deconstructing Anomaly Scores} \label{monitor}

Heatmaps provide a convenient way to visually monitor the anomaly scores and corresponding parameters over time.  In this section we display the standardized (z-score) shape signature parameters, ssr (sum of squared residuals), and anomaly scores over time; for two different ($tool$, $sensor$, $step$) triples, which we obfuscate as ``triple 1" and ``triple 2".  Each column in the heatmap gives the standardized parameter values of a different wafer; the wafers are ordered in time from left (earliest in time) to right (latest).   

\begin{figure}[h!]
\caption{Three shape signatures before and after the change point for triple 1. Curves corrected for x-shift, which does not contribute to anomaly score.  The black curve is the normal model.}
\includegraphics[
width=8.5cm]{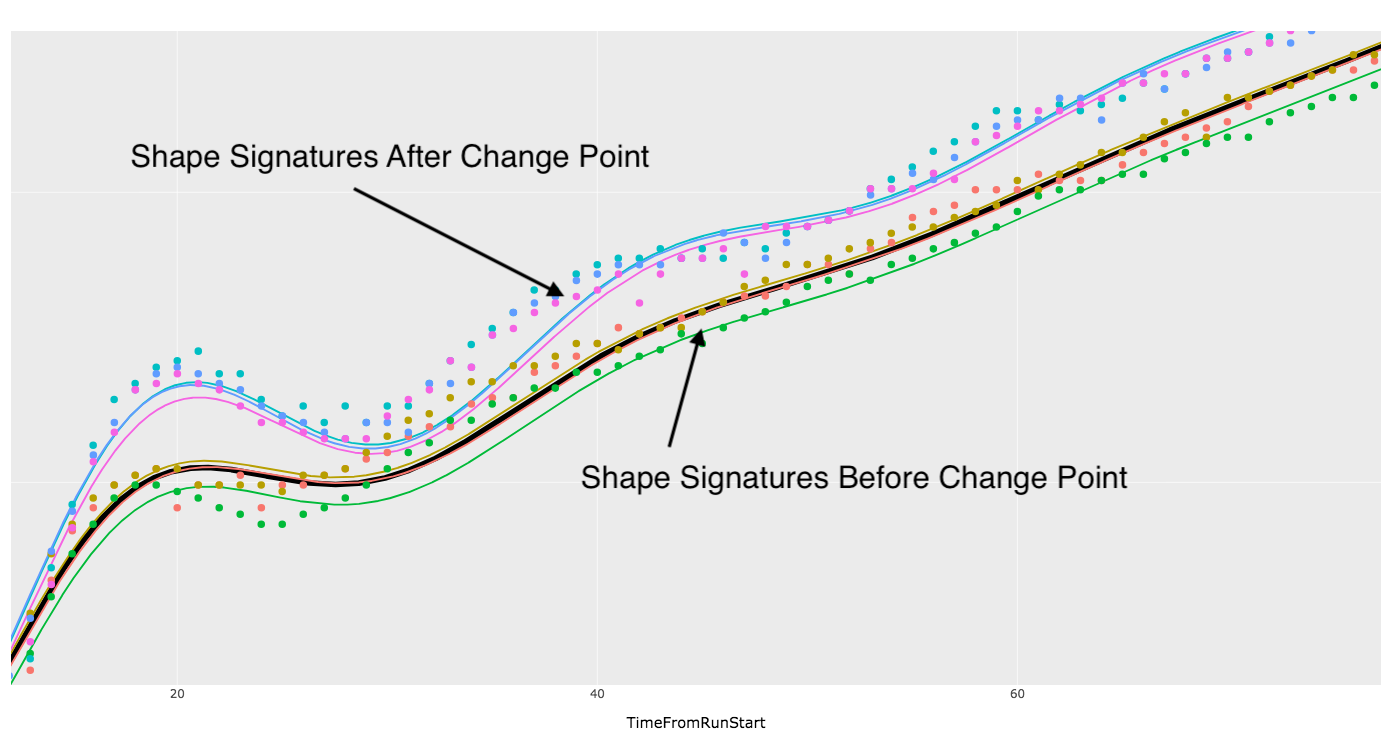}
\label{fig:2108A}
\end{figure}
\begin{figure*}[h!]
\caption{Time Evolution of Standardized Parameters for triple 1. Columns are wafers ordered by time.  We see almost all parameters shifting abruptly at the change point.}
\includegraphics[
width=18cm]{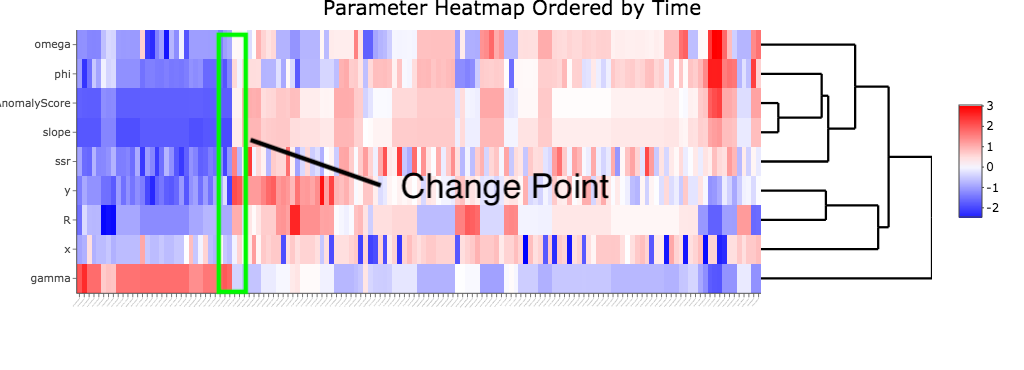}
\vspace{-0.5in}
\label{fig:heat2108A}
\end{figure*}

In Figure \ref{fig:heat2108A}, we see that the anomaly score and all parameters except for x (horizontal shift) experience an abrupt shift at the change point.  Note gamma changes in the negative direction and R, phi, and slope change in the positive direction.  Based on equation (\ref{diff sol}), these changes imply slower convergence to set point, larger amplitude, positive horizontal shift, and a steeper asymptote, respectively. These changes are apparent in Figure \ref{fig:2108A}, where we have plotted the shape signatures before and after the change point.  

We will now consider the gradient at the change point.  The change point is not readily available since it occurs in between two wafers.  We approximate the change point gradient by evaluating a first order Taylor series expansion of the gradient about the shape signature before the gradient at the midpoint of the shape signatures before and after the gradient.  Specifically, if $l_{change}$ is the wafer at which the change point occurs \footnote{Note that $l_{change}$ is a theoretical wafer, since it does not exist in our data.}, 
$l_{bef}$ is the available wafer immediately before the change point, and $l_{aft}$ is the available wafer immediately after the change point, then:
 
{\tiny
\begin{align*}
&\begin{pmatrix} 
\partial \text{anom}(s_{l_{change}})/\partial \gamma \\
\partial \text{anom}(s_{l_{change}})/\partial R \\
\partial \text{anom}(s_{l_{change}})/\partial \omega \\
\partial \text{anom}(s_{l_{change}})/\partial y \\
\partial \text{anom}(s_{l_{change}})/\partial \phi \\
\partial \text{anom}(s_{l_{change}})/\partial c \\
\partial \text{anom}(s_{l_{change}})/\partial x \\
\end{pmatrix} = 
\nabla \text{anom}(s_{l_{change}}) 
 \approx \nabla \text{anom}\left(\frac{s_{l_{bef}} + s_{l_{aft}}}{2}\right)
\\
& \approx \nabla \text{anom}(s_{l_{bef}}) + \nabla^2 \text{anom}(s_{l_{bef}})\left(\frac{s_{l_{aft}}-s_{l_{bef}}}{2}\right) \\
& = \begin{pmatrix}
-1.067120 \times 10^6 \\
4.223630 \times 10^3 \\
4.947775 \times 10^2 \\
7.326686 \times 10^1 \\
3.092527 \times 10^2 \\
1.623737 \times 10^6 \\
9.901231 \times 10^{-3}
\end{pmatrix}\ ,
\end{align*}
}%

Within this gradient, $\gamma$ and the slope $c$ most strongly contribute to the change in anomaly score, since they have the largest magnitude.  Less important contributors are $R$, $\omega$, and $\phi$; as their gradients are several orders of magnitude lower than those of $\gamma$ and $c$.  We see that $y$ and $x$ (vertical and horizontal shift, respectively) are not really contributing to the change point.  We also see that the signs of the elements of the gradient match up with what we observerd in figure \ref{fig:2108A}.  Specifically, $\gamma$ is contributing negatively and $c$, $R$, and $\phi$ are contributing positively.  Thus we have seen how the gradient is instrumental in establishing causal relationships between certain parameters and anomalous phenomena. 

\begin{figure*}[t!]
\caption{Time Evolution of Standardized Parameters for triple 2.  Columns are wafers ordered by time. The inset shows a zoom-in of the spike point. The anomaly score shoots up at the spike point and several other parameters also experience spikes correlated with anomaly score.}
\includegraphics[
width=18cm]{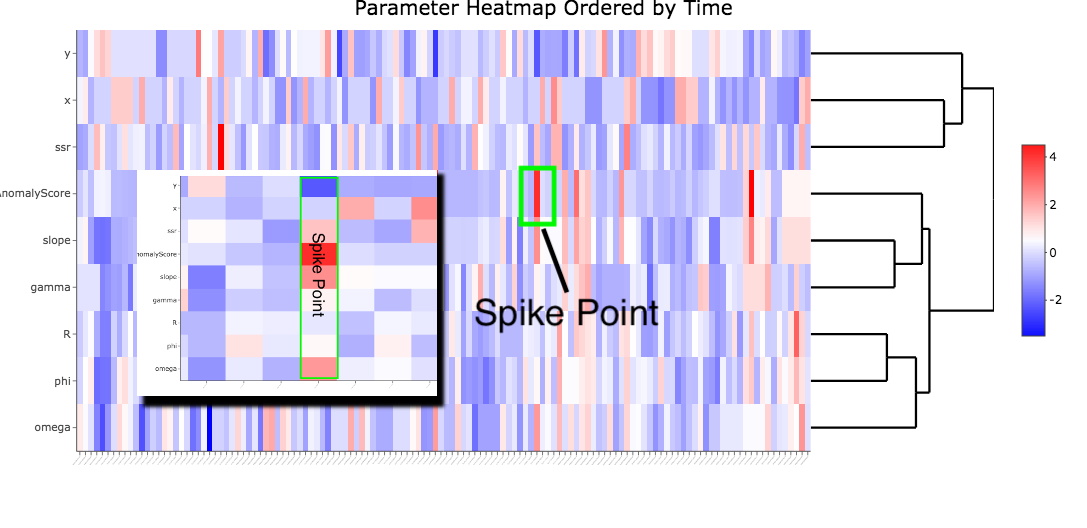}
\label{fig:heat2107A}
\vspace{-0.5in}
\end{figure*}

\begin{figure}[h!]
\caption{Triple 2 shape signatures in the neighborhood of the point where the anomaly score spikes.  The  downward shift in y (vertical shift) is clearly apparent. 
The black curve is the normal model.}
\label{fig:2107A}
\includegraphics[width=8.5cm]{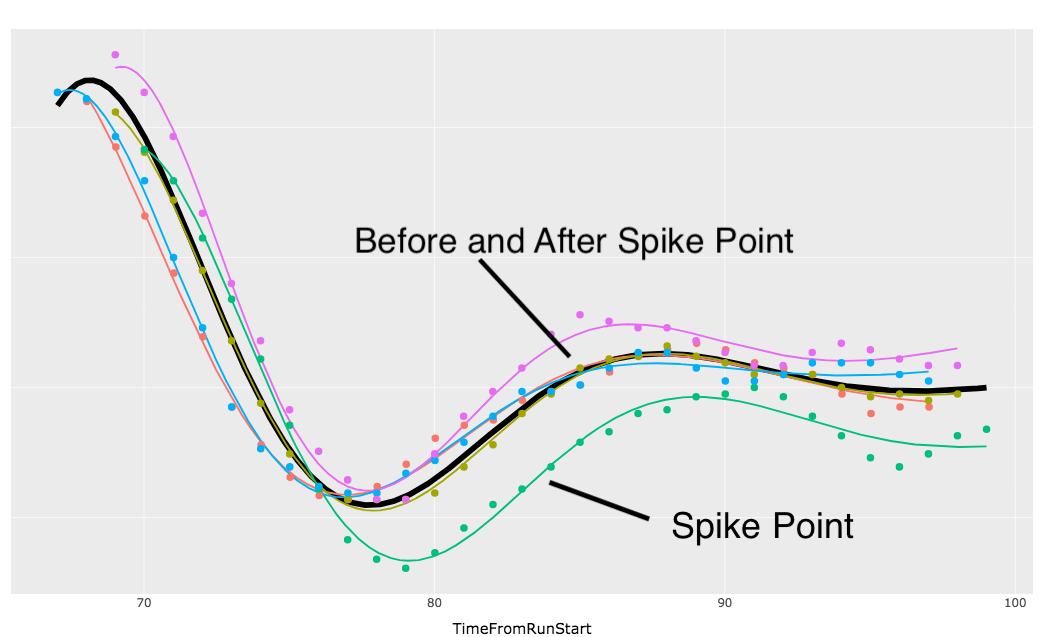}
\vspace{-0.25in}
\end{figure}

For ``triple 2", Figure \ref{fig:heat2107A} calls attention to a spike point, i.e. when anomaly score shoots up for just one wafer. Close inspection of the column of parameters at the spike point, shows   that the parameters ssr, slope, and omega are positively correlated with the anomaly score at the spike point; whereas y is negatively correlated (has a trough) with anomaly score at this point.  The trough in y is visually apparent in the raw data in Figure \ref{fig:2107A}, but the peaks of the other parameters are less perceptible by eye because these parameters vary on much smaller scales than y; they only stand out in the heatmap because of the standardization.  

By taking the gradient of the anomaly score, we can 
understand how changes in the parameters impact the anomaly score.  We expect $\partial \text{anom}(s)/\partial y$ to be negative at the spike point and the partial derivative with respect to the peaking variables to be positive at the spike point.  The gradient at $l_{spike}$, the wafer at which the spike occurred, confirms this:
$$
\small
\nabla \text{anom}(s_{l_{spike}})= 
\begin{pmatrix} 
\partial \text{anom}(s_{l_{spike}})/\partial \gamma \\
\partial \text{anom}(s_{l_{spike}})/\partial R \\
\partial \text{anom}(s_{l_{spike}})/\partial \omega \\
\partial \text{anom}(s_{l_{spike}})/\partial y \\
\partial \text{anom}(s_{l_{spike}})/\partial \phi \\
\partial \text{anom}(s_{l_{spike}})/\partial c \\
\partial \text{anom}(s_{l_{spike}})/\partial x \\
\end{pmatrix} = 
\begin{pmatrix} 
91.110112 \\ 
4.94883\\ 
169.607\\
-6.655819\\
10.89815\\
279336942 \\
0.0537148
\end{pmatrix}.$$

The gradient confirms the anomaly score correlations deduced
from the heatmap.  The magnitudes of the elements of the gradient give us the strength of the correlations, i.e. how strongly each variable is contributing to the spike.  The slope c is causing the spike point, as its corresponding partial derivative is significantly larger than all the rest.  In general, we can take the gradient of the anomaly score at points of interest to determine which shape signature parameters are influencing the anomaly score at that point.  

\section{Conclusion}


We have proposed a new ``grey-box" approach to anomaly detection in smart manufacturing of semiconductor wafers that combines feedback control theory and Bayesian statistical learning to discover causal relationships between deeply hidden control parameters and process behavior.   It is applicable to any manufacturing process, since these processes are known to be employ feedback loops.   

Our method fits shape signatures using a robust statistical model that features regularization and a second-order optimization routine.  The shape signatures provide powerful features, as each of their parameters have important physical meaning.  We kill two birds with one stone by finding a way to compute anomaly scores,  which automatically capture both goodness-of-fit and change in shape, using the same statistical model that we used to fit the shape signatures.   We illustrate ways in which these anomaly scores can be monitored over time using our semiconductor data.  The anomaly score can be decomposed into its individual components by taking its gradient.  Components of the gradient vector correspond to oscillator parameters that we have demonstrated how to map to the underlying feedback control parameters.  In future work, we can perform sensitivity analysis using perturbation theory using this mapping. 

Shape signatures provide a basis for powerful new methods for anomaly detection and deconstruction in smart manufacturing.  By explicitly modeling control systems,  they better capture normal system performance and provide deeper engineering insight.  This could potentially lead to higher manufacturing yields at lower costs.  The general approach of using parametric function representations of process control systems for quality control and process improvement can be adapted to other types of manufacturing processes and  control systems.

\section{Acknowledgments}
GLOBALFOUNDRIES provided partial funding and manufacturing process data. This work was  supported by NSF Grant 1331023, and the Rensselaer Institute for Data Exploration and Applications. We  thank Mr. Mark Reath of GLOBALFOUNDRIES for his support, mentoring and encouragement, and Dr. Wayne Baquette of Rensselaer Polytechnic Institute for his guidance on process control theory.

\bibliography{sigproc}

\begin{thebibliography}{1}

\bibitem{aggarwaloutlieranalysis}
C.~C. Aggarwal.
\newblock {\em Outlier Analysis}.
\newblock Springer International Publishing, Yorktown Heights, New York, USA,
  2017.

\bibitem{bequetteprocesscontrol}
B.~W. Bequette.
\newblock {\em Process Control: Modeling, Design, and Simulation}.
\newblock Pearson Education Inc., Upper Saddle River, NJ 07458, 2003.

\bibitem{cherry2006multiblock}
G.~A. Cherry and S.~J. Qin.
\newblock Multiblock principal component analysis based on a combined index for
  semiconductor fault detection and diagnosis.
\newblock {\em IEEE Transactions on semiconductor manufacturing},
  19(2):159--172, 2006.

\bibitem{ge2010semiconductor}
Z.~Ge and Z.~Song.
\newblock Semiconductor manufacturing process monitoring based on adaptive
  substatistical pca.
\newblock {\em IEEE Transactions on Semiconductor Manufacturing},
  23(1):99--108, 2010.

\bibitem{haq2016feature}
A.~A.~U. Haq, K.~Wang, and D.~Djurdjanovic.
\newblock Feature construction for dense inline data in semiconductor
  manufacturing processes.
\newblock {\em IFAC-PapersOnLine}, 49(28):274--279, 2016.

\bibitem{elementarydiffeqs}
R.~C.~D. William E.~Boyce.
\newblock {\em Elementary Differential Equations and Boundary Value Problems}.
\newblock John Wiley \& Sons, 111 River Street, Hoboken NJ 07030, 2012.

\bibitem{wise1999comparison}
B.~M. Wise, N.~B. Gallagher, S.~W. Butler, D.~D. White~Jr, and G.~G. Barna.
\newblock A comparison of principal component analysis, multiway principal
  component analysis, trilinear decomposition and parallel factor analysis for
  fault detection in a semiconductor etch process.
\newblock {\em Journal of Chemometrics: A Journal of the Chemometrics Society},
  13(3-4):379--396, 1999.

\end{thebibliography}
\nocite{bequetteprocesscontrol}
\nocite{aggarwaloutlieranalysis}
\nocite{elementarydiffeqs}
\bibliographystyle{abbrv}
\end{document}